# Quantum Bridge Analytics II: Network Optimization and Combinatorial Chaining for Asset Exchange

Fred Glover[1], Gary Kochenberger[2], Moses Ma[3] and Yu Du[4]

## Abstract

Quantum Bridge Analytics relates to methods and systems for hybrid classical-quantum computing, and is devoted to developing tools for bridging classical and quantum computing to gain the benefits of their alliance in the present and enable enhanced practical application of quantum computing in the future.

This is the second of a two-part tutorial that surveys key elements of Quantum Bridge Analytics and its applications. Part I focused on the Quadratic Unconstrained Binary Optimization (QUBO) model which is presently the most widely applied optimization model in the quantum computing area, and which unifies a rich variety of combinatorial optimization problems.

Part II (the present paper) examines an application that augments the use of QUBO models, by disclosing a context for coordinating QUBO solutions through a model we call the Asset Exchange Problem (AEP). Solutions to the AEP enable individuals or institutions to take fuller advantage of solutions to their QUBO models by exchanges of assets that benefit all participants. Such exchanges are generated by a combination of two optimization technologies, one grounded in network optimization and one based on a new metaheuristic optimization approach called combinatorial chaining. This combination provides a flexibility to solve AEP variants that open the door to additional links to quantum computing applications and additional applications via the Quantum Bridge Analytics perspective. We show how this modeling and solution capability gives rise to an *Asset Exchange Technology* that embraces a broad range of financial, industrial, scientific and social settings. Examples are presented that show the nature of these processes from a tutorial perspective.

[1]ECEE, College of Engineering and Applied Science, University of Colorado, Boulder, CO 80302 USA fred.glover@colorado.edu

[2,4]College of Business, University of Colorado at Denver, Denver, CO 80217 USA, gary.kochenberger@ucdenver.edu; yu.du@ucdenver.edu

[3]FutureLab Consulting, Mill Valley, CA, 94941 USA, moses.ma@futurelabconsulting.com

## 1. Introduction

Quantum Bridge Analytics is devoted to developing tools for bridging classical and quantum computing to gain the benefits of their alliance in the near term and enable enhanced practical application of quantum computing in the future.

As observed in Part I of this tutorial, the Quadratic Unconstrained Binary Optimization (QUBO) model has an important role in Quantum Bridge Analytics by unifying a rich variety of combinatorial optimization problems and becoming at present the most widely applied optimization model in the quantum computing area.

In Part II (the present paper) we examine an application that augments the use of QUBO models, by coordinating QUBO solutions through a model we call the *Asset Exchange Problem* (AEP), a problem class that likewise spans many important applications of optimization. Solutions to the AEP enable individuals or institutions to identify and profit from exchanges of assets that benefit all participants – exchanges that, in game theory terminology, constitute a positive sum game. This leverages the ability to take advantage of QUBO models by generating solutions either as inputs or outputs from these models to solve broader ranges of problems. Motivated by the Quantum Bridge Analytics perspective, we provide formulations and methods for the AEP model that we refer to by the title *Asset Exchange Technology*, which provides a set of tools enabling hybrid classical-quantum processes to yield positive sum games. As frequently observed in the context of economic exchange, such games are facilitated by the mechanisms of money, interest and middlemen. Asset Exchange Technology provides an additional mechanism for facilitating such exchange that serves the blockchain goal of disintermediation to remove or reduce reliance on intermediaries. The resulting modeling and solution process simultaneously afford a link between the applications of classical and quantum computing that are envisioned to be increasingly relevant as the quantum computing area becomes more mature.

We introduce two main hubs for Asset Exchange Technology, the first consisting of a mathematical formulation yielding a network optimization model for a basic version of the AEP and the second consisting of a metaheuristic optimization framework called combinatorial chaining that augments the network model to make it possible to derive high quality solutions to more complex instances of the AEP, notably including instances encountered in a wide variety of real world applications.

Asset Exchange Technology derives special relevance within the context of Quantum Bridge Analytics, which offers gains by bridging the gap between classical and quantum computational methods and technologies. As observed in the 2019 Consensus Study Report titled *Quantum Computing: Progress and Prospects* (National Academies, 2019), quantum computing will remain in its infancy for perhaps another decade, and in the interim "formulating an R&D program with the aim of developing commercial applications for near-term quantum computing is critical to the health of the field." The report further notes that such a program will rely on developing "hybrid classical-quantum techniques." Innovations that underlie and enable these hybrid classical-quantum techniques, which are the focus of Quantum Bridge Analytics, provide a fertile catalyst for introducing Asset Exchange Technology.

Additional links to the Quantum Bridge Analytics theme are provided in Kochenberger and Ma (2019) who observe that the QUBO model gives rise to a variety of formulations for portfolio optimization, and these in turn yield a natural basis for integrating classical and quantum computing via the Asset Exchange Problem. Portfolio optimization has a prominent role in the AEP when the assets under consideration involve those customarily incorporated into the portfolio domain. The AEP goes further, both in the portfolio domain and others, by linking the holders of multiple portfolios in a network of cooperative optimization. This establishes a natural alliance with QUBO models where QUBO solutions identify desirable assets for different participants and Asset Exchange Technology, notably via network optimization and combinatorial chaining, then solves AEP problems to find exchanges that meet disparate desirability criteria (embodied in these solutions) to benefit all members of each exchange.

The network and combinatorial chaining components of Asset Exchange Technology are joined within the framework of *netform* (network-related formulation) modeling, which concerns the development of models based on characterizing structure for the purpose of insight and more effective solution. As observed in Glover et al., (1990), the technologies of computer implementation and problem representation have profited from network optimization chiefly because advances in this field have intimately related problem solving to the identification and exploitation of structure.

We begin by introducing a general mathematical optimization model for a basic instance of the AEP and then show how the model can be transformed into a network optimization model, thus laying the foundation for exploiting more complex variants of the AEP via netform analysis. Combinatorial chaining provides the central mechanism for fulfilling this role by yielding a method to obtain high quality solutions for the basic problem that can readily be generalized to address more advanced variants.

A note on terminology: we use the term "exchange" rather than "swap" because a swap typically refers to an exchange involving only two items or two participants, and "multiple swaps" refer to a collection of pairwise exchanges, in contrast to an integrated process that requires the coordination among all participants for its execution.

The most developed literature on exchanges occurs for the traveling salesman problem, where the term *k-opt* refers to an exchange that removes k edges from a tour and replaces them by k other edges so that the resulting configuration continues to be a tour (Hamiltonian cycle; see, e.g., Helsgaun 2000, 2009). The traveling salesman procedures that come closest to the process of combinatorial chaining are the ejection chain approaches that have been applied to TSPs and other combinatorial optimization problems (Glover 1996; Rego and Glover 2006; Yagiura et al. 2006. 2007; Rego et al. 2016).

The blockchain literature refers to exchanges called *atomic swaps* (also known as cross-chain trading). As elaborated subsequently, these exchanges arise when two parties who want to share their cryptocurrencies execute an exchange by means of Hashed Timelock Contracts (or HTLCs) as a mechanism to make the transaction secure (Nolan 2013; Fitzpatrick 2019). Combinatorial chaining makes it possible to generalize these swaps to exchanges involving multiple actors.

Combinatorial chaining and the Asset Exchange Problem are to be differentiated from the problem and methods arising in combinatorial auctions where swaps are sought to exchange pairs of buy/sell-orders in futures markets (Winter et al. 2011; Müller et al. 2017). An interesting area for future investigation would be to determine if the combinatorial chaining approach could likewise be applied in the setting of combinatorial auctions to enable auctions involving greater numbers of participants.

The remainder of this paper is organized as follows. Section 2 gives examples of asset exchange applications to set the stage for later more extensive and technical discussions. Section 3 provides the fundamental mathematical formulation of the basic AEP problem, and shows how to transform this formulation into a network optimization model. Section 4 characterizes the structure of combinatorial chaining in reference to this basic network model, followed by introducing more advanced processes in Section 5 for joining network optimization and combinatorial chaining with metaheuristic analysis to address more complex instances of asset exchanges. The paper concludes with a summary of the key notions and their implications in Section 6.

## 2. Preliminary Examples of the Asset Exchange Problem

The Asset Exchange Problem (AEP) arises in a variety of contexts, spanning applications in financial investment, resource allocation, economic distribution and collaborative decision making. Our approach to solving this problem is based on a form of cooperative optimization, where multiple parties with complex criteria collaborate as well as compete for resources. This could apply to algorithms for distributing packages between trucks in a delivery network, or dynamic switching to alternative sorting facilities. Or it could apply to collaborative bidding processes for complex multi-criteria contracts or decentralized cooperative group optimization for multi-criteria investment cryptocurrency portfolios.

Cooperative group optimization provides a framework for implementing algorithm instances by integrating the advantages of the cooperative asset exchange in conjunction with low-level algorithm portfolio design. This is quite distinct from traditional portfolio optimization, as with a hedge fund that typically seeks to mitigate risk by diversification with some investments that are negatively correlated.

In the cooperative group optimization setting, our approach generalizes processes that seek atomic exchanges of baskets of fungible tokens or securities by yielding exchanges at a higher combinatorial level. Normally, a financial institution that wishes to execute a large basket of trades, in a way that mitigates execution risk by having an intermediary, can ask that institution to take the basket into its inventory and unwind the trades on its own. Thus, instead of revealing specific information about the assets in the basket, knowledge which could be exploited, the institution and banks can conduct a "zero-knowledge" protocol to effect basket trades. However, this protocol still requires trust in those institutions providing the service. The proposed new approach uses both simple and complex combinatorial exchanges to optimize all parties engaged in the multi-party optimization effort.

The progenitor of such an approach has emerged and is being tested in the cryptocurrency world – this is known as a *cross-chain atomic swap*. This is where two parties own tokens in separate cryptocurrencies, and want to exchange them without having to trust a third party or a centralized exchange. However, by extending this model and enabling complex multi-party exchanges, splits and aggregations, we can effect full spectrum combinatorial trading to provide *trustless algorithmic liquidity* without requiring even the normal underlying reserve trading currency.

The simplest instance of such a system is a marketplace of three portfolios. In this market, Portfolio A has asset X and wishes to own some asset Z, Portfolio B has asset Z but wishes to acquire only asset Y, and Portfolio C has asset Y and wishes to own some asset X. In a traditional exchange, participants would exchange what they have for the underlying reserve settlement currency, and then purchase what they want. This would entail two transaction fees per portfolio. Alternatively, using cross chain atomic swaps, the parties would never make any transactions whatsoever, as the global optimal cannot be reached via pairwise swap transactions.

By enabling all potential complex exchanges, splits and aggregations, for N portfolios, any market could increase its global utility. However, the computational complexity of this type of complex combinatorial exchange trading is NP-complete. By using a multi-attribute trade matching system that includes the unspoken goals of the parties, which are the "utility functions" of the parties, it is possible to find pareto-efficient exchange solutions – referring to the game theory concept of a strategy that cannot be made to perform better against one opposing strategy without performing less well against another.

Additionally, the inclusion of constraints increases the complexity of the problem. For example, if the system determines that diversification is required, then a constraint can be added that limits which types of assets could be included in the diversification target. Only assets that have been rated by a rating agency or analyst, for example, as better than a "B" rating, could be included to modify the optimization. A continuous approach would assign numerical value a rating, and blend that with volatility metrics, volume data, social impact scores, and even the user's personal pet peeves – to enable a multi-objective approach to optimize both individual and multiple portfolios.

In the future, the user will require the ability to enter or modify both market orders (with fixed prices) and limit orders (with variable prices). As we transition from market orders to limit orders, this will help to expand utility expression, and it can become appropriate to add constraints to help identify price improvement opportunities – allowing a combinatorial exchange to operate for a share of price improvement, rather than charging transaction fees. Just a Bitcoin promises "zero cost transactions", this could provide a model for "zero cost exchanges" that provide the appearance of negative transaction cost given a disparity of utility functions. In section 5 we discuss the use of priorities to address such considerations.

The current model for the most effective form of exchange is the double-sided exchange, a system in which both buyers and sellers provide bids for matching via the exchange. A central controlling system matches the sell bids with the buy bids, yielding matched buy bids and matched sell bids in response thereto so that allocations of the matched buy bids and the matched sell bids maximize the throughput of the exchange. Combinatorial exchanges using cooperative optimization could potentially lay a foundation for the next evolutionary step in market exchange protocols, moving from double sided trading using a reserve currency to something more general that encompasses new forms of economic transactions.

Double sided exchanges are used to trade goods, services, or other things of value, including network bandwidth trading, financial-instruments trading, transportation logistics, pollution-credit trading, electric power allocation, and so on. However, to make double sided exchanges work, they require fungibility. And so, varying levels of quality, that describe for example the quality of crude oil, are lumped into fungible categories of sulfur content, gravity, etc. This further suggests the possibility that combinatorial exchanges could reflect multi-attribute trading more effectively, allowing traders to work with greater accuracy in pricing.

Combinatorial exchanges can likewise be used for handling non-fungible assets. As long as people are willing to assign value to objects to be traded, combinatorial exchanges can provide a basis to get people what they want. Suppose User A wants to sell a vacation timeshare he or she is tired of, for a certain collectible car with roughly the same value. There are no matches as both are relatively illiquid markets and it could take several months or require a significant discount to find buyers. However, there could be a User B who has exactly the car A wants, but doesn't want a timeshare, and instead wants a diamond necklace. Now if there is a jeweler C who would find that timeshare exciting, and willing to create a custom necklace to B's liking, the system could enable algorithmic liquidity by joining all three into a complex transaction.

Moving toward a more general example, A and B's assets most likely have different values. If there is no jeweler willing to make just the right necklace, the value exchange would not add up. Two parties would likely need to add or accept part of the value in cash. However, with the inclusion of a special user D, who is willing to inject cash and accept a partial tokenized share of that collectible car or real estate, the complex transaction become possible. We call this special user a "decentralized market maker" who would require a modest premium to compensate for enabling greater liquidity. That token share could be sold at a later time, hence it is an offer to sell cash for time.

Decentralized market making is an intriguing concept that would require a detailed exploration, as it will likely emerge as critical factor for enabling scalable liquidity. But there are many questions to be answered. For example, what is the value of contributions by the decentralized market makers? Also, could these small investments held by the market – provided to equalize values in an exchange –be aggregated into baskets, and could those baskets be traded? How do we accurately assess the risks of items in baskets, to flow them up to the basket, to avoid "toxic assets" being included?

Finally, it should be noted that a computational system or agent that learns what a user wants to buy or sell, or might be willing to trade, would be quite valuable as an e-commerce tool because it provides a means to unveil the deeper purchase intentions of users. AI based agents could assist not only in the process of helping the user to determine what they might be willing to trade for or buy but could even help the user discover new purchase intentions that might lead to greater personal satisfaction. In other words, instead of just contributing to the accumulation of more useless stuff in their lives, such a system could explore more complex human values, as opposed to those reflecting desires and whims stimulated by media and advertising.

For example, if an AI held a model that understood the OCEAN Big Five personality traits, which was used so effectively by Cambridge Analytica in 2016, it could predict that the user has a high degree in a single trait, openness to new experiences. By balancing knowledge around both investment planning and personality traits, the advisor could provide more balanced advice to the user that would lead to greater personal satisfaction and fulfillment. A strictly financial based AI advisor would simply recommend one asset class over other, or the diversification into additional classes. But an AI advisor that used both financial optimization as well as heuristics about human personality and psyche, understanding the complex needs of the investor... might suggest to keep 95% of the portfolio within financial instruments, but propose that 5% could be invested in experiential learning for the user, in other words, investing in him or herself. This could include travel to learn a new language or a workshop to learn a new skill, possibly with permission to tap into the user's online "bucket list" – the list of things you'd like to do before you "kick the bucket."

To put this into the context of the AEP problem and combinatorial chaining, consider a situation with User A who has inherited a somewhat odd abstract painting from a distant relative in France, that doesn't have much value on the resale market in America. However, on a combinatorial exchange market, there may be a chance of trading it for something not only less objectionable but desirable for all parties. Her asking price is a value of $3,000. Now, because her interaction with the exchange is managed by a user agent with access to her private "bucket list," the trusted agent can now look for something that matches items on his list. It turns out that she has always wanted to take a class at the Cordon Bleu cooking school and to learn some French. So our agent can scan against other agents and listings, to find User B who wants to trade a $3000 workshop pass at Cordon Bleu for ten day stay in a beachfront Airbnb on some nice tropical island. The combinatorial chain holds that in place while finding a third or fourth transaction to make the combinatorial exchange pareto-optimal for all users. Fortunately, it finds User C who has a modest bungalow on a beach in the Marquesas, which doesn't get much Airbnb interest because it is too remote. However, that person looks at the painting, and realizes it was painted by the singer Jacques Brel, who was a great singer but lousy painter, and actually has quite a bit of value in the Marquesas because Jacques Brel spent his last days on the island of Hiva Oa, following the footsteps of Paul Gauguin and learning how to paint untamed landscapes that were so bad they looked abstract. So his agent offers a 3 week stay for that painting!

In this way, an AI-based financial advisor would advise in a more human and humane way. Thus, metaheuristic optimization via asset exchange technology could be applied directly to the issues of happiness, life goals and meaning. For user A, the lifelong goal of learning how to master the art of French cooking. For User B, a desperately needed vacation he couldn't afford otherwise. And for User C, the lifelong goal of appearing on Antique Roadshow, to show off a barn find of a lifetime. We thus can ascend from cold process of optimizing utility functions to optimizing the human condition.

One last note concerns the potential for quantum computing in this arena. In general, present day quantum computers can handle only a very limited number of qubits, representing a small number of asset types, or cooperating portfolios. When quantum computers can offer tens or hundreds of thousands of qubits, with effective partitioning algorithms, combinatorial exchanges will be able to scale to manage real world liquidity needs for applications involving massive numbers of participants and classes of items to exchange. Until then, quantum computing will enable exchange functionality for only limited and constrained markets, such as for cryptocurrencies. For example, a crypto wallet that holds only a dozen types of crypto would represent a relatively small variable space and could potentially be optimized using a quantum computer. Money was invented to simplify barter, and a quantum exchange based on pareto-efficient combinatorial exchanges could simplify money.

Motivated by the Quantum Bridge Analytics perspective we can go beyond the present limitations of quantum computing to provide these exchanges by identifying combinatorial chaining algorithms that are capable of accommodating variable spaces for AEP models of significantly greater dimension, providing advances in the near term that can be translated into progressively greater advances in the future as quantum computing technology becomes more mature.

### 3. Mathematical Formulations of the AEP

The Asset Exchange Problem has several levels. We start from the most basic level of the AEP in reference to a graph $G = (N, E)$, with node set $N = \{1, \ldots, n\}$ and edge set $E = \{\{i,j\}, i, j \in N\} \subset N \times N$. Each node $i \in N$ identifies an entity such as an individual or business or institution, and each edge $\{i,j\}$ identifies an *exchange link* between i and j. Let *A* denote the set of asset types (classes), where elements $\alpha \in A$ can represent classes of tokens in a cryptocurrency application or types of securities in a securities market or categories of commodities in a commodity market, and so forth. In the following we use the term *assets* interchangeably with the term *asset types*.

Each node $i \in N$ has a set $S_i$ of assets it can send (i.e., can agree to send) to other nodes and a set $R_i$ of assets it can receive (i.e., can agree to receive) from other nodes. Thus, for example, if $\alpha' \in S_i$ and $\alpha'' \in R_i$, then node i can agree to send asset $\alpha'$ and agree to receive asset $\alpha''$ from other nodes. More precisely, $R_i$ denotes assets that i desires (considers beneficial) to receive and $S_i$ denotes assets that i is willing to send (in return for obtaining an asset in the set $R_j$). We say a transfer of asset $\alpha$ from node i to node j is *admissible* if $\alpha \in S_i$ and $\alpha \in R_j$ ($\alpha \in S_i \cap R_j$). We allow only admissible transfers in seeking asset exchanges that benefit all participants.

Define $N_i = \{j \in N: \{i,j\} \in E\}$ to be the set of nodes j that are neighbors of node, i.e., that join node i by an edge. Let $x_{ij}^\alpha$ denote the number of units of asset $\alpha$ transferred from node i to node j. In restricting consideration to admissible transfers, we assume each node i has an upper limit $U_i^{\alpha:R}$ on the number of units of any given asset $\alpha \in R_i$ that can be admissibly be transferred from other nodes to node i and an upper limit $U_i^{\alpha:S}$ on the number of units of $\alpha \in S_i$ that can be transferred from i to other nodes. Formally, these conditions may be expressed as

$$\sum(x_{ij}^\alpha: j \in N_i) \leq U_i^{\alpha:R} \quad i \in N \text{ and } \alpha \in S_j \cap R_i \quad (1)$$
$$\sum(x_{ij}^\alpha: j \in N_i) \leq U_i^{\alpha:S} \quad i \in N \text{ and } \alpha \in S_i \cap R_j \quad (2)$$

We also impose an equation that requires the total number of assets transferred from a given node i to other nodes j to equal the total number of assets transferred in return from other nodes j to node i. Specifically, for a given node $i \in N$ and a given node $j \in N_i$, we observe that the quantity $\sum(x_{ij}^\alpha: \alpha \in S_i \cap R_j)$ identifies the total number of units that can be admissibly transferred from node i to node j and similarly, the quantity $\sum(x_{ij}^\alpha: \alpha \in S_j \cap R_i)$ identifies the total number of units that can be admissibly transferred from node j to node i. We require these two quantities to be equal by stipulating

$$\sum(x_{ij}^\alpha: \alpha \in S_i \cap R_j) = \sum(x_{ij}^\alpha: \alpha \in S_j \cap R_i) \quad i \in N \text{ and } j \in N_i \tag{3}$$

Finally, we impose an additional limit $U_i$ on the number of all assets $\alpha$ that can be admissibly transferred from node i to other nodes, expressed as

$$\sum(x_{ij}^\alpha: j \in N_i, \alpha \in S_j \cap R_j) \leq U_i \quad i \in N \tag{4}$$

As a result of equation (3), this inequality is equivalent to

$$\sum(x_{ij}^\alpha: j \in N_i, \alpha \in S_j \cap R_i) \leq U_i \quad i \in N \tag{4'}$$

Subject to these conditions, in the AEP we seek to maximize the total number of admissible exchanges, hence yielding the complete formulation

$$\text{Maximize} \quad \sum(x_{ij}^\alpha: i \in N, j \in N_i, \alpha \in S_i \cap R_j) \tag{0}$$
subject to (1), (2), (3), (4) and $x_{ij}^\alpha \geq 0$, $i \in N$, $j \in N_i$ and $\alpha \in S_i \cap R_j$

We can also replace (0) by a variety of other objectives, such as

$$\text{Maximize} \quad \sum(p_i^\alpha x_{ij}^\alpha: i \in N, j \in N_i, \alpha \in S_i \cap R_j) \tag{0'}$$

where $p_i^\alpha$ is a positive monetary value that node i attaches to receiving asset $\alpha$ from the set $R_i$.

We now take the step of transforming the preceding formulation into a network optimization formulation, to give a foundation for generating solutions to the foregoing AEP model by a corresponding basic version of our combinatorial chaining approach. From this, we will be able to treat related more complex models by natural extensions that combine the network optimization and combinatorial chaining components. The transformation to a network formulation significantly increases the problem size, but offsets this by making the problem sparser, while our combinatorial chaining algorithm for this formulation is able to work with a memory based on the number of nodes rather than the number of arcs in the network, dramatically reducing both the amount of computation and the memory involved.

**The Network AEP Formulation**

The AEP network formulation arises by replacing the graph G by a graph $G^* = G^*(N^*, A^*)$ consisting of a set of nodes $N^*$ and a set of arcs (directed edges) $A^*$ as follows.

To emphasize the arc orientation in creating G*, we find it useful to augment the customary representation of an arc from a node p to a node q as an ordered pair (p, q) by alternatively writing it in the form p → q, which adds clarity when p and/or q is itself represented as an ordered pair. Lower bounds on all arc flows are assumed to be 0.

We divide each node $i \in N$ into two nodes, i[R] and i[S], and create an arc i[R] → i[S], with an upper bound on its flow of $U_i$ from (4). Then for each $i \in N$ and $\alpha \in R_i$ we create new nodes ($\alpha$, i[R]), producing $\sum(|R_i|: i \in N)$ nodes, and create arcs ($\alpha$, i[R]) → i[R] (from node ($\alpha$, i[R]) to node i[R]) which results in $\sum(|R_i|: i \in N)$ arcs (the same as the number of nodes ($\alpha$, i[R]). Each of these arcs receives the upper bound $U_i^{\alpha:R}$ from (1) to limit its flow. Similarly, for each $i \in N$ and $\alpha \in S_i$ we create new nodes ($\alpha$, i[S]), producing $\sum(|S_i|: i \in N)$ nodes, and create arcs i[S] → ($\alpha$, i[S]), creating $\sum(|S_i|: i \in N)$ arcs (the same as the number of nodes ($\alpha$, i[S]). Each of these arcs receives the upper bound $U_i^{\alpha:S}$ from (2) to limit its flow. It is assumed that $U_i$ satisfies $U_i \leq \text{Min}(\sum(U_i^{\alpha:R}: \alpha \in R_i), \sum(U_i^{\alpha:S}: \alpha \in S_i))$, that is, the upper bound $U_i$ on the flow across arc i[R] → i[S] is limited by the smaller of the sum of upper bounds on the arcs ($\alpha$, i[R]) → i[R] entering i[R] and the sum of upper bounds on the arcs i[S] → ($\alpha$, i[S]) leaving i[S]. (Later we also describe variations in which we additionally introduce lower bounds $L_i^{\alpha:R}$ and/or $L_i^{\alpha:S}$ on the arcs ($\alpha$, i[R]) → i[R] and arcs i[S] → ($\alpha$, i[S]).)

Finally, for each $i \in N$ and for each $j \in N_i$ such that $\alpha \in S_i$ is the same as $\alpha \in R_j$ (i.e., for which $\alpha \in S_i \cap R_j$), each node ($\alpha$, i[S]) joins by an arc ($\alpha$, i[S]) → ($\alpha$, j[R]). We call these the *α-linking arcs* of G*, since the same asset $\alpha$ is referenced by both nodes of each of these arcs. The number of these arcs is $\sum |S_i \cap R_j|: i \in N, j \in N_i)$.

From this construction we see that N* consists of $2n + \sum(|R_i| + |S_i|: i \in N)$ nodes and A* contains $n + \sum(|R_i| + |S_i|: i \in N) + \sum |S_i \cap R_j|: i \in N, j \in N_i)$ arcs.

Because we start from the symmetric graph G in undirected edges to produce the graph G* with directed arcs, we assume $j \in N_i$ implies $i \in N_j$. We additionally observe that no asset $\alpha$ is contained in both $R_i$ and $S_i$ for any given i, under the assumption that if node i sees a benefit in receiving a unit of $\alpha \in R_i$, then it will not be willing to relinquish a unit of $\alpha$ by including it in $S_i$. Exceptions can be imagined, as where i may be willing to give up a particular $\alpha' \in R_i$ if it is able to receive a more highly valued asset $\alpha'' \in R_i$. Such exceptions can be modeled by extensions of the constructions used here but make the formulation larger and more complex. Nevertheless, our basic algorithm can be modified to handle these and other variations without entailing the complexity introduced by an extended mathematical formulation.

The foregoing description of G* can be translated into an algorithm for generating the graph. As part of this we show how to attach numerical indexes denoted by k = 1 to n* to the nodes in N* so that G* may be represented as a network in a standard format. We refer to lower bounds as well as upper bounds on arcs for generality, although in most circumstances lower bounds will be 0.

**Algorithm to Generate G***

For each i ∈ N
   Create the nodes i[R] and i[S] and the arc i[R] → i[S], by assigning the index k = i to the
      node i[R] and the indexes k = i + n to the node i[S].
   Attach the lower and upper bounds $L_i$ and $U_i$ to the arc i[R] → i[S]  (i → i + n).
Endfor
Set k = i + n
For each i ∈ N
   (Create the "S-labeled" asset node (α, i[S]) and associated arc i[S] → (α, i[S]) for each
      asset α ∈ $S_i$.)
   For each α ∈ $S_i$
      Set k: = k + 1 and create the asset node (α, i[S]), assigning it the index k.
      Create the arc i[S] → (α, i[S]) and attach lower and upper bounds $L_i^{\alpha:S}$ and $U_i^{\alpha:S}$.
   Endfor
Endfor
For each j ∈ N
   (Create the "R-labeled" asset node (α, j[R]) and associated arc (α, j[R]) → j[R] for each
      asset α ∈ $R_j$.)
   For each α ∈ $R_j$
      Set k: = k + 1 and create the asset node (α, j[R]), assigning it the index k.
      Create the arc (α, j[R]) → j[R] and attach lower and upper bounds $L_j^{\alpha:R}$ and $U_j^{\alpha:R}$.
   Endfor
Endfor
For each i ∈ N
   (Create the "S to R" asset arcs (α, i[S]) → (α, j[R]) associated with i for each α ∈ $S_i$.)
   For each asset α ∈ $S_i$
      For each neighbor j ∈ $N_i$
         For each asset α ∈ $R_j$ such that α ∈ $R_j$
            Create the asset arc (α, i[S]) → (α, j[R]) with no bounds (i.e., a lower
               bound of 0 and an upper bound of infinity).
         Endfor
      Endfor
   Endfor
Endfor

Costs or profits may be attached to the arcs of the network according to the objective that is desired to be achieved. Asset arcs, which are linking arcs, should be assigned a 0 cost or profit. In creating asset arcs (α, i[S]) → (α, j[R]) above, if there is no asset α ∈ $R_j$ for any j ∈ $N_i$ such that α ∈ $S_i$, then the asset node (α, i[S]) will not have any corresponding node (α, j[R]) to create such an arc (α, i[S]) → (α, j[R]), and α can be dropped from $S_i$. If all α ∈ $S_i$ are thus removed to leave $S_i$ empty, then i can be removed from N and i[R] and i[S] (= i and i + n) correspondingly removed from N*. Similarly, if at the end it is discovered that the node (α, j[R]) has no arcs (α, i[S]) → (α, j[R]) entering it, then this implies that α can be dropped from $R_j$. By the same token, if all α ∈ $R_j$ are thus removed to leave $R_j$ empty, then j can be removed from N and j[R] and j[S] (= j and j + n) correspondingly removed from N*.

An illustration of the graph G* is given in Appendix 1.

### 4. Basic Version of Combinatorial Chaining

A classical theorem of network flows (Fulkerson and Ford, 1962) implies that a feasible solution to the network formulation of the AEP can be decomposed into a collection of cycles (not necessarily disjoint or uniquely determined). Such cycles are of interest for the Asset Exchange Problem because they identify a collection of participants who can enter into a succession of mutually beneficial asset exchanges. Such a collection is not unduly difficult to identify by reference to a solution to the AEP network formulation but requires additional effort. More importantly, a standard network flow algorithm for solving the basic AEP formulation is not capable of being directly adapted to provide good solutions to more complex variations of the AEP that abound in practical applications, thus motivating the creation of the adaptive combinatorial chaining approach.

Adopting the netform perspective (Glover et al., 1992), combinatorial chaining is designed both to exploit the structure of the basic AEP network formulation and to be susceptible to extensions for solving a variety of AEP variations found in practice. This harmonizes with the Quantum Bridge Analytics perspective as in applications where quantum computing can be applied to solve portfolio optimization problems expressed as QUBO models for individual investors or institutions and combinatorial chaining can then be applied to the appropriate AEP variation to integrate and improve these individual solutions to the benefit of each participant.

The strategy underlying combinatorial chaining operates by generating successions of directed trees (or arborescences in graph theory) rooted at different nodes. Conditions are monitored to disclose when a directed tree can be extended by connecting a tip of one of its branches to the root, thus creating a cycle that constitutes a mutually beneficial exchange. The process differs from classical tree generation algorithms by introducing multiple categories of tree predecessors and establishing a mechanism to trace the predecessors that differentiates between the categories effectively. This departure from classical approaches arises because the AEP belongs to the class of multi-commodity network flow problems (Hu, 1963; Assad, 1978), which are more complex than standard "pure" network flow problems, and normally cannot be expressed as a pure network problem as we have accomplished for the AEP. Rather than being a disadvantage, however, this complexity fosters a response that enables the chaining mechanism to be adapted to other AEP variations.

**Combinatorial Chaining for the Basic Network AEP**

Combinatorial chaining for the basic network AEP makes use of arrays denoted FlowR(α, i[R]) to the record flows on the arcs (α, i[R]) → i[R] and arrays denoted FlowS(α, i[S]) to record the flows on the arcs i[S] → (α, i[S]). Hence, for each i ∈ N, we require FlowR(α, i[R]) ≤ $U_i^{α:R}$ for each $\alpha \in R_i$, and require $FlowS(\alpha, i[S]) \leq U_i^{\alpha:S}$ for each $\alpha \in S_i$. Flows on the arcs arc $i[R] \rightarrow i[S]$ are recorded in an array $Flow(i)$ for each $i \in N$. All flow values are initialized to 0.

It is convenient to refer to the nodes $(\alpha, i[R])$, $(\alpha, i[S])$ and i (the latter collectively representing the two nodes $i[R]$ and $i[S]$) as *open* when their associated flows $FlowR(\alpha, i[R])$, $FlowS(\alpha, i[S])$ and $Flow(i)$ do not reach their upper bounds and *closed* otherwise. (A bit can be set for each such node to determine its open/closed status.)

We refer to two types of predecessor arrays $PredR(i)$ and $PredS(i)$, $i \in N$, accompanied by associated arrays $AssetR(i)$ and $AssetS(i)$ explained subsequently. The arrays $PredR(i)$ and $PredS(i)$ are initialized to 0 to indicate predecessors are not yet assigned.

The method performs forward scans and reverse scans to examine nodes $i \in N$ (and from there to examine the arcs these nodes can become linked to in a chain). When a tip of the tree can successfully be linked to the root, a *breakthrough* occurs by establishing the existence of an exchange cycle that is mutually beneficial for all its node i participants. Breakthrough is accompanied by appropriately updating (increasing) the flows on arcs of the cycle.

The basic version of the chaining algorithm only performs forward scans but gives the foundation for performing reverse scans as well, as subsequently described. We first explain the nature of the forward scan routine and then give a more formal description.

*Rationale of the Forward Scan Routine*:
The Forward Scan Routine is embedded in a Main Routine that maintains a set $N^o$ identifying the open nodes, initialized by $N^o = N$. Nodes to be scanned are placed in a set denoted ScanSet that begins with a chosen node $i^* \in N^o$. During the Forward Scan Routine, ScanSet acquires other nodes $i \in N^o$ to form a tree that yields a collection of chains rooted at node $i^*$. The tree is generated by successively selecting new nodes i from ScanSet as long as ScanSet $\neq \emptyset$.

For each node i selected from ScanSet, consider each asset $\alpha \in S_i$; i.e., each asset $\alpha$ that node i is willing to send to another node. Given node i, additionally consider each neighbor j of i that contains $\alpha$ in $R_j$; i.e., each neighbor j that desires to receive $\alpha$. (Formally, we refer to the set $NR_i^\alpha = \{j \in N_i: \alpha \in R_j\}$, which consists of those neighbors j of node i such that $R_j$ contains $\alpha$.) If node j is not already in the tree, i.e., if it has no predecessor (as indicated by $PredS(j) = 0$), then it can acceptably be added to the tree by adopting node i as its predecessor. For this, we set $PredS(j) = i$ together with $AssetS(j) = \alpha$, which records the fact that each chain in the tree that passes through this particular (i. j) link is accompanied by sending asset $\alpha$ from node i to node j.

If now $j = i^*$ (which can result because $i^*$ is not assigned a predecessor initially), we have discovered a chain beginning with node $i^*$ that results in a loop which qualifies as a mutually beneficial exchange cycle (where each participant receives a desired asset and in return sends a willingly exchanged asset). The Breakthrough Routine handles this outcome by identifying the cycle and updating the flows and the structure of G* appropriately.

Following the updates of the Breakthrough Routine, the scanning routine is reinitiated within the Main Routine by selecting a new i* from $N^o$ (where i* may be the same as before if it is not removed from $N^o$ during breakthrough).

Alternatively, the scan from a given node i* may terminate with ScanSet empty and without achieving breakthrough. In this case, i* is removed from $N^o$ and once more the scanning routine is reinitiated within the Main Routine to select a new i* from $N^o$.

We let $N_i^o = N_i \cap N^o$ denote the (current) neighbors of node i that are in $N^o$. Hence $N_i^o$, which starts the same as $N_i$, may shrink as nodes are removed from $N^o$. This also modifies the definition $NR_i^\alpha = \{j \in N_i: \alpha \in R_j\}$ to become $NR_i^\alpha = \{j \in N_i^o: \alpha \in R_j\}$, identifying the neighbors of i in $N^o$ that desire to receive asset $\alpha$.

Termination of the Main Routine occurs when $N^o$ contains only a single node ($|N^o| = 1$), since then this node has no other nodes it can exchange with.

The formal design of the algorithm is as follows.

***Combinatorial Chaining Algorithm***
*Initialization.*
Set all flow values to 0. Initialize the set $N^o$ of open nodes by setting $N^o = N$.
*Main Routine*
While $|N^o| > 1$
    Set all predecessor arrays to 0.
    Choose i* $\in N^o$ and create ScanSet = {i*}.
    Execute the *Forward Scan Routine* (as follows)
    While ScanSet $\neq \varnothing$
        Select a node i $\in$ ScanSet
        For each $\alpha \in S_i$
            For each $j \in NR_i^\alpha$ ($= \{j \in N_i^o: \alpha \in R_j\}$)
                If PredS(j) = 0 then
                    (j has not been visited before on a Forward Scan)
                    Set PredS(j) = i and AssetS(j) = $\alpha$.
                  If j = i* then
                      Execute the *Breakthrough Routine* (below)
                      (Update flows and potentially remove nodes from $N^o$.)
                      Break (leave Forward Scan Routine to choose a new i* $\in N^o$ in the
                      Main Routine if $|N^o| > 1$).
                Endif
            Else
                Let ScanSet := ScanSet $\cup$ {j}.
                Endif
            EndFor
        EndFor
        ScanSet = ScanSet\{i} (remove i from ScanSet)
        (The scan of node i is complete.)

EndWhile
        (End of the Forward Scan Routine)
Endwhile
(End of the Main Routine)

The algorithm can be modified to save part of the tree after the completion of each forward scan, but the computational savings will not usually be enough to warrant the effort. Reverse scanning provides a more interesting modification and can be accomplished by interchanging R and S in each of the instructions of the Forward Scanning Routine. Forward scanning and reverse scanning can also be done together, switching from one to the other on selected iterations. In this case, breakthrough is recognized when j = PredS(i) on a forward scan yields PredR(j) > 0 (where PredR(j) was set on a reverse scan), or when j = PredR(i) on a reverse scan yields PredS(j) > 0 (where PredS(j) was set on a forward scan). To show how reverse scanning can be joined with forward scanning, Appendix 2 gives an example where a single iteration of reverse scanning is applied before launching the forward scanning algorithm.

The Breakthrough Routine that accompanies the Forward Scanning Routine may now be described as follows. The preceding observations and the example in Appendix 2 disclose how to modify this routine for reverse scanning or for combinations of forward and reverse scanning.

***Breakthrough Routine***
*Compute the maximum feasible flow increment ΔFlow on the augmenting cycle*
ΔFlow = Big (a large positive number)
i = i*
Stop = False
While Stop = False
    α = AssetS(i)
    ΔR = $U_i^{\alpha:R}$ – FlowR(α, i[R])
    i = PredS(i)
    ΔS = $U_i^{\alpha:S}$ – FlowS(α, i[S])
    Δi = $U_i$ – Flow(i)
    ΔFlow = Min(ΔR, ΔS, Δi, ΔFlow)
    If i = i* then Stop = True
Endwhile
*Update flows and remove nodes associated with saturated arcs*
Let ε denote a small positive number (to provide tolerance for roundoff error)
i = i*
Stop = False
While Stop = False
    α = AssetS(i)
    FlowR(α, i[R]) = FlowR(α, i[R]) – ΔFlow
    If FlowR(α, i[R]) > $U_i^{\alpha:R}$ – ε  then close arc (α, i[R]) by setting $R_i := R_i \setminus \{\alpha\}$
        (removing α from $R_i$)
    i = PredS(i)
    FlowS(α, i[S]) = FlowS(α, i[S]) – ΔFlow
    If FlowS(α, i[S]) > $U_i^{\alpha:S}$ – ε  then close arc (α, i[S]) by setting $S_i := S_i \setminus \{\alpha\}$

```
        (removing α from S_i)
    Flow(i) = Flow(i) − ΔFlow
        If Flow(i) >  U_i − ε  then close arc (i[R], i[S]) setting Nº := Nº \{i}
    If i = i* then Stop = True
Endwhile
```

## 5. Extensions and Metaheuristic Connections

There are problems that are too complex to be given mathematical formulations that fully capture their subtleties and that are simultaneously capable of being solved by standard math programming algorithms. In adopting the perspective of Quantum Bridge Analytics, we embrace strategies for such problems that allow their objectives to be pursued approximately and flexibly, thus admitting approaches that solve variations of these problems to emphasize alternative problem components in an adaptive fashion. As we have emphasized, our combinatorial chaining procedure allows this to be done when joined with network optimization by giving a framework that yields access to more complex variants.

We show how this can be achieved for two chief extensions of the basic formulation that encompass a broad range of applications. The associated modified versions of combinatorial chaining provide flexible approximation methods that can be embedded in metaheuristic algorithms and afford the possibility of being incorporated into hybrid classical/quantum systems. In common with the most effective algorithms for QUBO problems, a natural basis for these extensions derives from adaptive memory strategies such as embodied in tabu search and path relinking (Glover and Laguna, 1997; Wang et al., 2012; Samorani et al., 2019).

*Prioritizing the assets exchanged*

In some applications of the AEP, participants may wish to prioritize certain exchanges of assets over others, preferring more strongly to receive particular assets and being more willing to relinquish certain other assets. Priorities attached to these preferences may also differ among different participants. Upon assigning numerical values to capture these preferences (as by indicating a dollar amount that different individuals attach to the value of different exchanges, or by making recourse to an agreed-upon set of subjective weights), the combinatorial chaining algorithm can be extended by prioritizing the selection of the elements i* in Nº or the choice of elements i in ScanSet, in each instance selecting the highest priority element from those available.

Priorities can also be used by such an extension to improve the choices for participants whose exchanges were less favorable on previous executions of the algorithm, since an effort to achieve a best overall collection of exchanges (such as a maximum number of beneficial exchanges) can result in better outcomes for some participants than for others. This means of exploiting the freedom to choose different elements in executing the basic steps of combinatorial chaining yields an approximation method for a problem whose subtleties render it unsuitable for a classical mathematical formulation, while allowing the flexibility to be adapted to different types of priorities. Such priorities can be introduced in the network formulation as and embodied in probabilities for selecting moves in metaheuristic adaptations as in probabilistic tabu search (Xu et al., 1997; Guermi et al., 2019). Combinatorial chaining provides the underlying structure for guiding the search to produce feasible solutions.

Priorities can also be employed to create larger breakthroughs earlier in the process of generating combinatorial chains, as by giving higher priority to participants with larger capacities (upper bounds) on the flows they can receive. The priorities can be based on measures applied to each base node (participant), such as total sums of capacities or means of capacities adjusted by standard deviations, and so forth. Refinements arise by considering the priorities of neighbors. For example, a new priority can be created for a node that is a weighted combination of its current priority and the priorities of neighbor nodes, where weights for neighbors are less than for the node under consideration. Such a process may also be repeated, using the new priorities as a basis for constructing another round of new priorities. (Additional repetitions may be expected to yield progressively less advantage.)

Particular applications give their own criteria for determining priorities. In exchanges of cryptocurrencies, for example, larger investors face the most negative impact by failing to make exchanges of a size deemed satisfactory, so assigning higher priorities to exchanges of such investors will usually result in the highest increase in utility. Using such priorities, choosing a node i* from $N^o$ with the highest priority to become the root of the current directed tree, followed by choosing highest priority nodes i from ScanSet to continue building the tree, provides a compelling and easily implemented strategy.

As previously observed, there may also be situations where it can be relevant to place lower bounds as well as upper bounds on the number of units of different assets exchanged by different participants. In a cryptocurrency application, for example, an investor may only be interested in transactions that result in receiving a specified number of units of a given asset. To illustrate, an investor represented by a node i may seek an exchange in which i receives precisely 100 units of Ethereum (ETH), represented by asset α ($\in R_i$) (accompanied, for example, by i sending units of Bitcoin (BTC) or Lumen (XLM) to other nodes). The AEP network model then captures this by putting a lower bound of 100 and an upper bound of 100 on the ETC arc (α, i[R]) → i[R]), giving $L_i^{\alpha:R} = U_i^{\alpha:R} = 100$. The situation where an investor may have an exact demand for an asset (modeled by setting the lower bound equal to the upper bound), and where this demand cannot be satisfied by an exchange involving any single other investor, is sometimes called *splitting*, i.e., the demand must be split into different transactions with different investors. Combinatorial chaining automatically handles splitting situations as well as other much more general situations. A simple illustration is where investor i will only consider an exchange that brings in at least $L_i^{\alpha:R} = 50$ units of ETH, but would prefer to receive more units, up to a limit of $U_i^{\alpha:R} = 100$. Any number of other investors, some who may not be neighbors of i, may be involved in transactions identified by combinatorial chaining.

In cases like these where the AEP model includes lower bounds on numbers of units received, exchanges can be prioritized in two phases, where Phase 1 is devoted to satisfying as many of the lower bounds as possible, and Phase 2 then sends additional flow through the network subject to satisfying upper bounds. These two phases are not required to have the same priorities for selecting nodes on exchange cycles.

Machine learning provides a natural way to facilitate priority generation. A strategy of varying the priorities may yield better overall outcomes for a particular objective, for example, and machine learning can be used to help identify a strategy that leads to the most desirable results. An instance of machine learning called Programming by Optimization (Hoos, 2012) is often effective for choosing parameters for optimization algorithms and may be useful in determining priorities in the combinatorial chaining context. Learning can also be employed as clustering-based metaheuristics (Samorani, et al., 2019).

*Generalized networks*

An important extension of the AEP arises where a unit of one asset may be exchanged for more or less than one unit of another asset. Networks in which the number of units received at the destination node (to-node) of an arc may differ from the number of units sent from the origin node (from-node) of an arc are called *generalized networks*, (Glover et al., 1990, 1992) and the factor that determines the difference between the units sent and received is called the *arc multiplier*. For example, an arc multiplier of 1.5 implies that the to-node receives 1.5 units for every unit sent from the from-node. A variety of situations exist where assets may be exchanged other than on a one-to-one basis.

A convenient feature of the basic combinatorial chaining algorithm is that such multiplier effects can be captured by joining the treatment of priorities with a modification of the Breakthrough Routine. The amount of flow transmitted across a chain of generalized arcs on a path leading from the root node $i^*$ to a subsequent node $i$ equals the product of the multipliers on the arcs between $i^*$ and $i$. Thus, for example, if the chain consists of the succession of arcs $(i^*, i1)$, $(i1,i2)$, $(i2,i3)$, with $i3 = i$, and if the multipliers on these three arcs are 0.6, 2.0 and 1.2, then a unit of flow sent from node $i^*$ becomes $0.6 \times 2.0 \times 1.2 = 1.44$ units of flow received at node $i3$. The Breakthrough Routine can be readily modified to incorporate this effect, using it to identify the limits on flows required to compute updated flows across the entire cycle and to determine which assets or elements must be removed from their associated sets due to these updates.

The approaches of introducing exchange priorities and capitalizing on the ability to incorporate arc multipliers in association with generalized networks can be combined to cover an additionally expanded range of practical problems, which may be usefully exploited by metaheuristic algorithms in the QBA context.

## 6. Concluding Remarks

The relevance of Quantum Bridge Analytics for real world applications has been demonstrated by showing an important instance where we are able to apply the QBA perspective to the challenging Asset Exchange Problem, which opens up numerous applications in financial investment, resource allocation, economic distribution and collaborative decision making. The linkage of network optimization with metaheuristic optimization via combinatorial chaining gives rise to an Asset Exchange Technology that can address and solve a wide range of practical variations.

Present day quantum computers can only handle small AEP problems, due to the limited number of qubits they encompass, but the integration of network formulations and combinatorial chaining is capable of accommodating AEP problems of significantly greater dimension. Through these connections, the AEP model gives an important class of optimization problems that can be usefully approached within the QBA domain, providing a foundation for progressively greater advances in the future as quantum computing technology becomes more mature.

## Appendix 1: Illustration of Network Structure

The structure of the network created in Section 3 is illustrated in the following diagram, where the *i nodes* are represented in their duplicated form i[R] and i[S], giving rise to the arc i[R] → i[S], for a network with N = {1, …, 6}. The assets α are represented by the letters A, B, C, D and E, giving rise to *asset nodes* of the form (X, i[S]) and (X, j[R]) which are joined by arcs (X, i[S]) → (X, j[R]) (called α-linking arcs in Section 3), where i and j may vary but the asset X (= A, B, …, etc.) must be the same in each such arc. It should be noted that these linking arcs do not have limiting bounds on their flows other than an implicit lower bound of 0.

The arcs of the network that can be represented by a succession of columns of R-labeled nodes and S-labeled nodes, in a pattern that begins with the R-labeled i nodes i[R], followed by the S-labeled i nodes i[S], followed in turn by the S-labeled asset nodes (X, i[S]), then followed by the R-labeled asset nodes (X, j[R]) and finally followed by the R-labeled i nodes i[R] to repeat the pattern. A further interesting pattern seen in the diagram is that all S-labeled nodes have exactly 1 arc entering but may have multiple arcs leaving, while all R-labeled nodes have exactly 1 arc leaving but may have multiple arcs entering. The i nodes are enclosed in circles in the diagram and the asset nodes are enclosed in rectangles.

# NETWORK STRUCTURE

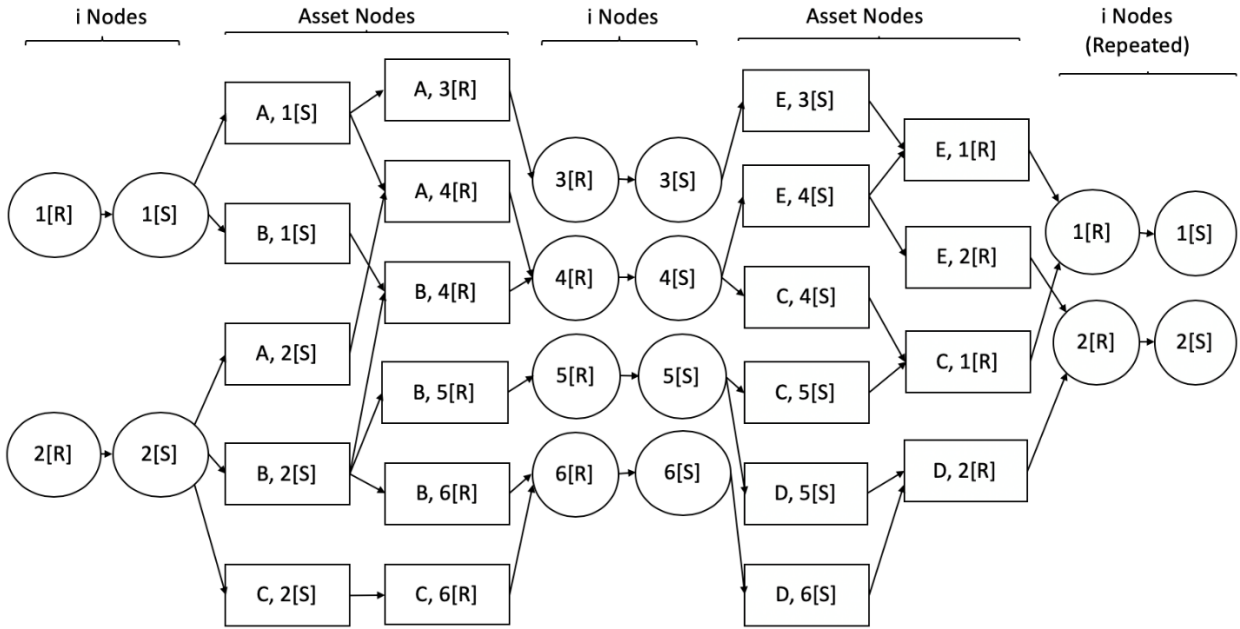

Since the asset arcs (linking arcs) do not have bounds on their flows, the foregoing pattern implies that an asset arc whose S-labeled node has a single arc out can be collapsed to be represented only by the R-labeled node, and an asset arc whose R-labeled node has a single arc in can be collapsed to be represented only by the S-labeled node. It should be emphasized that the staged structure shown in the diagram above is slightly misleading, since cycles typically vary in length and, in addition, duplicated i nodes may be encountered at various stages without implying they form a cycle that can be traced back to a previous instance of a duplicated node. The i indexes and the assets in the diagram have been ordered to show the patterns produced by arranging the nodes in columns. By contrast, the algorithm given in Section 3 for generating the network applies for any ordering of the indexes i in N and is independent of any ordering of the assets, which shows that such orderings are irrelevant in the general case.

# Appendix 2: Illustration for Reverse Scanning

*Combinatorial Chaining Algorithm with a Reverse Scanning Step*
*Initialization.*
Set all flow values and all predecessor arrays to 0. Initialize the set $N^o$ of open nodes by setting $N^o = N$.

*Main Routine*
While $|N^o| > 1$
    Set all predecessor arrays to 0.
    Choose $i^* \in N^o$ and create ScanSet = $\{i^*\}$.
    Execute the *Reverse Scan Routine* (below) to identify Fertile nodes $j \in N^o$
        (recorded by setting PredR(j) = $i^*$).
        If no Fertile nodes are found (Find = False), set $N^o := N^o \setminus \{i^*\}$ and Continue the next
        iteration of the Main Routine (returning to choose a new $i^* \in N^o$ (accompanied by
        ScanSet = $\{i^*\}$) if $|N^o| > 1$).
    Execute the *Forward Scan Routine* (as follows)
    While ScanSet $\neq \emptyset$
        Select a node $i \in$ ScanSet
        For each $\alpha \in S_i$
            For each $j \in NR_i^\alpha$ $(= \{j \in N_i^o: \alpha \in R_j\})$
                If PredS(j) = 0 then
                    (j has not been visited before on a Forward Scan)
                    Set PredS(j) = i and AssetS(j) = $\alpha$.
                    If PredR(j) > 0, then (node j is a Fertile node)
                        Execute the *Breakthrough Routine* (below)
                        (Update flows and potentially remove nodes from $N^o$.)
                        Break (leave Forward Scan Routine to choose a new $i^* \in N^o$ in the
                        Main Routine if $|N^o| > 1$).
                Endif
            Else
                Let ScanSet := ScanSet $\cup \{j\}$.
            Endif
            EndFor
        EndFor
    EndWhile
    (End of the Forward Scan Routine)
Endwhile
(End of the Main Routine)

*Reverse Scan Routine*
Set Find = False
For each $\alpha \in R_{i^*}$
    For each $j \in NS_{i^*}^\alpha$ $(= \{j \in N_{i^*}^o: \alpha \in S_j\})$

   If PredR(j) = 0 then
    (j has not been visited before on this Reverse Scan)
    Set PredR(j) = i* and AssetR(j) = $\alpha$ and Find = True
   Endif
  Endfor
Endfor
If (Find = False) then no fertile nodes are discovered.

*Note*: Find = False at the end only if $NS_{i*}^{\alpha} = \emptyset$ for all $\alpha \in R_{i*}$. The check for PredR(j) can be ignored and the assignments PredR(j) = i* and AssetR(j) = $\alpha$ and Find = True can be executed for each j encountered. It doesn't matter that these assignments write over previous assignments in this case.